\documentclass[12pt]{emulateapj}
\usepackage{epsfig}

\def\spose#1{\hbox to 0pt{#1\hss}}

\def\multleft#1{\hbox to size{\vbox {\halign {\lft{##}\cr #1}}\hfill}\par}
\def\multright#1{\hbox to size{\vbox {\halign {\rt{##}\cr #1}}\hfill}\par}

\def\boxit#1{\vbox{\hrule\hbox{\vrule\kern3pt\vbox{\kern3pt
          #1 \kern3pt}\kern3pt\vrule}\hrule}}

\def\beq{ \begin{equation} }
\def\eeq{ \end{equation} }

\newcommand\bb[1] {   \mbox{\boldmath{$#1$}}  }

\newcommand\del{\bb{\nabla}}

\def\gtsima{$\; \buildrel > \over \sim \;$}
\def\gtsim{\lower.5ex\hbox{\gtsima}}

\newcommand\bcdot{\bb{\cdot}}

\newcommand\dd{\partial}



\begin{document}

\title{Regulation of thermal conductivity in hot galaxy clusters by MHD turbulence}

\author{Steven~A.~Balbus\altaffilmark{1} and
Christopher~S.~Reynolds\altaffilmark{2}}

\altaffiltext{1}{Laboratoire de~Radioastronomie, \'Ecole Normale
Superi\'eure, 24 rue Lhomond, 75231 Paris CEDEX 05}
\altaffiltext{2}{Department of Astronomy, University of Maryland,
College Park, MD 20742-2421}

\begin{abstract}
The role of thermal conduction in regulating the thermal behavior of
cooling flows in galaxy clusters is reexamined.  Recent investigations
have shown that the anisotropic Coulomb heat flux caused by a
magnetic field in a dilute plasma drives a dynamical instability.
A long standing problem of cooling flow theory has been to understand
how thermal conduction can offset radiative core losses without
completely preventing them.  In this {\em Letter}, we propose that
magnetohydrodynamic turbulence driven by the heat flux instability
regulates field-line insulation and drives a reverse convective thermal
flux, both of which may mediate the stabilization of the cooling cores of
hot clusters.  This model suggests that turbulent mixing should accompany
strong thermal gradients in cooling flows.  This prediction seems to be
supported by the spatial distribution of metals in the central galaxies
of clusters, which shows a much stronger correlation with the ambient
hot gas temperature gradient than with the parent stellar population.

\end{abstract}

\keywords{cooling flows --- galaxies: clusters: general --- magnetic fields
--- MHD --- instabilities --- turbulence}

\section{Introduction}

Although the intracluster medium (ICM) of galaxy clusters has been an
active area of study since the earliest days of X-ray astronomy
(Felten 1966), the complexity of the physical processes regulating the
behavior of this gas continues to surprise us.  X-ray observations
imply that cooling times in the core regions of ICM atmospheres are
significantly shorter than cluster ages (Peterson \& Fabian 2006).
Prior to the launch of {\it XMM-Newton} and {\it Chandra}, it was
widely held that these short cooling times lead to flows in which mass
is dropping out (Fabian 1994).  Subsequent {\it XMM-Newton}
spectroscopy confirmed the short cooling times and inwardly decreasing
ICM temperature profiles, but also revealed a {\it lack} of gas
cooling below temperatures of $T\sim T_{\rm vir}/3$, where $T_{\rm
vir}$ is the virial temperature of the cluster (Peterson et al. 2001;
Tamura et al. 2001).  This observation, combined with a realization
that unchecked cooling flows would lead to central galaxies far above
the high-end cutoff of the luminosity function (Benson et al. 2003),
made it clear that the {\it cooling cores} of galaxy clusters must be
continuously heated.

A new paradigm consisting of core heating by a central radio galaxy
has taken hold.  Although a few examples were known at the time of
{\it ROSAT,} {\it Chandra} and {\it XMM-Newton} have observed cavities
and weak shocks in the ICM of many clusters that are driven by the
central radio-galaxy (e.g., Fabian et al. 2000, 2003; Young et
al. 2002; Heinz et al. 2002).  This has motivated the hypothesis that
core losses are offset by thermalization of the kinetic luminosity of
the jets from the active galactic nucleus (AGN).  In some (but not
all; see \S4) systems, the energetics of the AGN is comparable to the
radiative luminosity of the cluster core (Birzan et al. 2004; Best et
al. 2007).  AGN feedback of this form has been adopted by
semi-analytic galaxy formation theories in the guise of {\it
radio-mode} feedback (Croton et al. 2006).

An alternative hypothesis is that thermal conduction into the core
from the large heat reservoir beyond the cooling radius may offset the
radiative loss (Binney \& Cowie 1981).  A source of uncertainty for
such models is the factor $f$ by which magnetic fields suppress
thermal conduction below the classical Spitzer value $\kappa_S$
(Spitzer 1962).  Plausible estimates range from $f\sim 10^{-2}$
(Churazov et al. 2001) to $f\sim 0.3$ (Narayan \& Medvedev 2001).
Thermal conduction models have faced other significant challenges,
however.  Even unbridled Spitzer conductivity cannot offset radiative
cooling in the lower-mass/cooler clusters (Voigt and Fabian 2004), a
consequence of the strong temperature dependence of Spitzer
conductivity, $\kappa_S\propto T^{5/2}$.  For hotter clusters, thermal
conduction can in principle offset radiative cooling if $f\sim 0.1-1$
(Zakamska \& Narayan 2003).  In this case, however, fine tuning of $f$
is required to offset radiative cooling, yet not erase the observed
temperature gradients within the cooling core (e.g., see Conroy \&
Ostriker 2008).  Furthermore, assuming $f$ is a constant for a given
ICM atmosphere, these equilibria are still thermally unstable (Kim \&
Narayan 2003).

In this {\it Letter}, we argue that recent theoretical developments
in the theory of the stability of MHD atmospheres must have a crucial
bearing on the behavior of cluster cooling cores.  In particular, it is
now understood that in atmospheres of dilute plasmas (such as the ICM)
in which thermal conduction is restricted to follow magnetic field
lines, the convective instability criterion is fundamentally altered:
temperature gradients, not entropy gradients, are crucial.  Atmospheres
in which the temperature gradient is decreasing upwards ($dT/dz<0$), but
field lines are isothermal, are convectively unstable (Balbus 2000, 2001;
Parrish \& Stone 2005, 2007) even when the entropy gradient would normally
imply stability.  On the other hand, atmospheres in which the temperature
gradient is increasing upwards ($dT/dz>0$; the case in cooling cores), and
field lines are thermally conducting, are {\it also} unstable (Quataert
2008; Parrish \& Quataert 2008).  These two powerful instabilities are
known respectively as the magnetothermal instability (MTI; $dT/dz<0$)
and the heat flux-driven buoyancy instability (HBI; $dT/dz>0$).

Our focus is the cooling core of a rich/hot cluster of galaxies.  We
investigate the consequences of convective turbulence that is
driven, but also {\it regulated}, by the HBI.  Turbulence, far from
being a source of heat, is actually a cooling term in the energy
balance.  The role of convective fluctuations is an outwardly-directed
``return heat flux'' that, together with field line insulation,
lowers, but by no means erases, the inward diffusive thermal flux.
This process eliminates the theoretical {\em ad hoc} fine tuning
required of previous conduction-based models.  Since our picture
attributes the stabilization of hot cooling cores to the intrinsic
dynamics of MHD plasmas, the observed 
%
%
%similarity 
%
homology of hot cooling cores range is perhaps not surprising.
Moreover, turbulence-regulated heat conduction also accounts for the
otherwise puzzling tight correlation between metalicity distribution
and temperature gradient observed in cluster cores: the two become
intimately related to the presence of convective turbulence.

An outline of this {\em Letter} is as follows.  In \S2, we discuss the
role of radiative losses on the HBI in cluster cooling flows.  
%Our
%fundamental physical model is developed in \S3.  We contrast our
%picture with the AGN feedback model in \S4, and summarize our
%conclusions in \S5.
The role of convective heat transport is studied in \S3, and our
picture for cooling cores is explicitly discussed in \S4.  We
summarize our conclusions in \S5.

\section{Radiative cooling effects on the HBI}

Before presenting our thermal model of hot cluster cooling cores, it
is of interest to modify the HBI dispersion relation of Quataert
(2008) due to the presence of radiative losses.  We restrict our
calculation to the simplest and most illustrative case of a uniform
background magnetic field in the vertical $z$ direction.

Radiative losses are represented by the net cooling function
$-\rho{\cal L}$ (units erg s$^{-1}$ cm$^{-3}$);
this terms includes any explicit heating processes as well as
radiative cooling.  
We make the formal assumption that the background
is static.  To include first-order perturbations in radiative loses,
the term $-\delta(\rho{\cal L})$ should be added to the right side of
equation (11) in Q08.  
Plane wave disturbances of the form $\exp(\sigma t+ik_x x+ik_z z)$
then satisfy the dispersion relation
\beq\label{C}
\sigma_{therm}\sigma^2_{dyn} +{k_x^2\over k^2} \sigma N^2 
-{2\over 5}\kappa {k^2_z k^2_x\over k^2} g {d\ln T\over dz} =0
\eeq
where
\beq\label{B}
\sigma_{therm}= \sigma +{2\over5}\kappa k_z^2 +\left[{2T\over 5P}
{\left( \dd(\rho{\cal L})\over \dd T\right)_P}\right]
\eeq
\beq\label{A}
\sigma^2_{dyn} = \sigma^2 + (\bb{k\cdot v_A})^2 
\eeq
We have written $\sigma$ for $-i\omega$ in Q08, otherwise the notation is
identical.  (The quantity $g$ is $-(1/\rho){dP/dz}$.)  

For the case of a vertical field, equation (\ref{C}) corresponds
precisely to the dispersion relation (13) of Q08, but with the 
additional cooling term in square brackets in equation (\ref{B}). 
The constant term in the dispersion relation (\ref{C}) is
\beq
(\bb{k\cdot v_A})^2
\left[{2\over5}\kappa k_z^2 +{2T\over 5P}
\left( \dd(\rho{\cal L})\over \dd T\right)_P\right]
-{2\over 5}\kappa {k^2_z k^2_x\over k^2} g {d\ln T\over dz}
\eeq
This should be negative for instability.  In the limit of very small
$\bb{k\cdot v_A}$, we recover the HBI condition that an increasing
temperature profile is unstable.
%
%
%But the presence of a magnetic field
%in principle allows a positive temperature gradient to be present and
%marginally stable at wavelengths comparable to the size of the region.
%Thermal cooling is stabilized by the conduction term $(2/5)\kappa k_z^2$,
%but at longer wavelengths it is possible for
%radiative thermal instability to significantly strengthen the HBI.
%
We see, however, that in the presence of a non-negligible magnetic
field, radiative cooling (which, in this setting, has $\left(
\dd(\rho{\cal L})/\dd T\right)_P<0$) acts to further destabilize
the atmosphere, especially at longer wavelengths.

The non-linear behavior of an HBI unstable atmosphere is still
unclear.  On the basis of 2D and 3D simulations of non-cooling
plane-parallel atmospheres, Parrish \& Quataert (2008; hereafter PQ08)
show that the HBI acts to re-orient field lines perpendicular to the
temperature gradient, effectively shutting off the conductive heat
flux.  We shall refer to this process as {\it field-line insulation}.
However, it is not obvious that the conclusions of PQ08 can be
extended to actual cluster cooling cores.  As well as the topological
impossibility of establishing a (non-zero) field configuration that
has zero radial field everywhere, the inflow of matter associated with
unbalanced radiative cooling in an insulated core will readily
generate a radial field from existing transverse field.  Thus, we
argue that field-line insulation, while surely present at some level,
cannot be absolute in cluster cooling cores.

One of the most salient properties of the linear HBI is its convective
nature, in particular the fact that the product $\delta v_z\, \delta T
>0$ (Q08).  If sufficiently large, this reverse flux 
regulates the background heat flux: the greater the magnitude
of $dT/dz$, the more vigorous the return flux.  We suggest that
the combination of field line insulation and reverse convective
flux fixes a state of marginal stability to the radiative HBI
that corresponds to the cool cores of galaxy clusters.

\section{Heat transport within hot cluster cooling cores}

Let us now re-consider the energy equation for the cooling core of a hot
cluster in the light of the above discussion.  Under steady
conditions, the equation of hydrodynamical energy conservation is
\beq
\del\bcdot \left[
\left(
{v^2\over2}+{5P\over 2\rho} +\Phi\right)\rho\bb{v}
+\bb{Q}\right] =-\rho{\cal L},
\eeq
where $\Phi$ is the potential function of the cluster, and
$\bb{Q}=\kappa_S\bb{b}(\bb{b}\cdot\del T)$ is the diffusive heat
flux ($\bb{b}$ is a unit vector in the direction of the magnetic 
field).

We wish to investigate the nature of dynamically {\it static}
solutions to this equation in which, on average, the mass flux
$\rho\bb{v}$ vanishes.  The HBI is assumed to be present and fully
developed into convective turbulence.  This requires a background heat
flux $\bb{Q}$ to be present, since this is the seat of the
instability.  We assume that the density and temperature consist of a
mean background value plus a fluctuation, $\rho = \rho_0 +\delta\rho,
\quad T=T_0+\delta T$.
%
%
%\beq
%\rho = \rho_0 +\delta\rho, \quad T=T_0+\delta T
%\eeq
%
The radial velocity $v$ (the only component of concern here)
is dominated by its fluctuating component, but must also
have a very small second order part with a finite mean value,
$v_2$.  This is needed if the mass flux vanishes,
\beq
\langle (\rho_0 +\delta\rho)(v_2+\delta v)
\rangle \simeq \rho_0v_2 + \langle \delta\rho\,
\delta v\rangle = 0
\eeq
where $\langle\rangle$ indicates a time average.  Defining 
the normalized temperature $\tau=P/\rho$, our energy equation 
reads
\beq
\del\bcdot\left[{5\over 2} \rho_o \langle\delta \bb{v}\, \delta \tau\rangle
+\bb{Q}\right] = -\rho\cal{L}
\eeq
%
%
%This explicitly demonstrates that it is the divergence of both
%the (inwardly directed) diffusive heat flux $\bb{Q}$ and the
%(outwardly-directed) turbulent heat flux $(5/2)\rho \langle\delta
%\bb{v}\delta \tau\rangle$ that, in our model, balances radiative
%cooling.
%
%Notice that the form of the convective heat flux is proportional to
%the product of a velocity and temperature correlation.  By contrast,
%Parrish \& Stone (2007) and Parrish \& Quataert (2008) both estimated
%the flux by using expressions proportional to the product of velocity
%fluctuation and the pressure, which yields a considerably smaller estimte
%of the convective flux.  We believe the velocity-temperature correlation
%should be used for the convective flux.
%
%We now estimate the outwardly directed convective
%heat flux $(5/2)\rho \langle\delta \bb{v}\delta \tau\rangle$.
%(Henceforth, we drop the $0$ subscript: $\rho$ and $T$ are understood
%to be background values.)  The argument of Schwarzschild (1958) for
%stellar convection together with the results in Q08 can be 
%adapted to the problem at hand.  The fastest growing modes of the HBI
%typically have (Q08)
%
This explicitly demonstrates that it is the divergence of both the
(inwardly directed) diffusive heat flux $\bb{Q}$ and the
(outwardly-directed) turbulent heat flux $(5/2)\rho \langle\delta
\bb{v}\delta \tau\rangle$ that, in our model, balances radiative
cooling.  Notice that, as in Schwarzschild (1958), the convective
heat flux is proportional to the correlation of velocity and temperature
fluctuations. 

We now estimate the outwardly directed convective heat flux $(5/2)\rho
\langle\delta \bb{v}\delta \tau\rangle$, using arguments similar to
Schwarzschild (1958).  Henceforth, we drop the $0$ subscript: $\rho$
and $T$ are understood to be background values.  The fastest growing
modes of the HBI typically have (Q08)
\beq
{\delta \tau\over \tau} = {\delta T \over T} = \xi {d\ln T\over dr},
\eeq
and density deficits 
\beq
| \delta \rho | = \rho \xi {d\ln T\over dr},
\eeq
where $\xi$ is the radial displacement.  Following Schwarzschild, we
multiply this density deficit by $g\xi/2$
%
%
%, where $g$ is the
%gravitational acceleration, 
to obtain the kinetic energy per unit volume of a rising element,
$\rho(\delta v)^2/2$:

\beq
{\rho (\delta v)^2\over 2} = {\rho g\over 2} \xi^2 \left(d\ln T\over dr\right),
\eeq
or
\beq
\delta v =  \xi  g^{1/2} \left(d\ln T\over dr\right)^{1/2}
\eeq
Therefore, the convective heat flux is
\beq\label{chf}
{5\over 2}\rho \langle\delta v \delta \tau\rangle = {5\over 2}P g^{1/2}
\left(d\ln T\over dr\right)^{3/2} \xi^2 = {5\over2}P\left(d\ln T\over
dr\right){\cal D},
\eeq
where ${\cal D}$ is the turbulent diffusion coefficient
\beq\label{D}
{\cal D}\equiv
\langle \xi \delta v\rangle =\xi^2 g^{1/2} \left(d\ln T\over dr\right)^{1/2}.
\eeq

Schwarzschild identifies $\xi$ with $l/2$, where $l$ is a formal
mixing length, but such details are of secondary concern here.  The
main point is that if these basic scalings are correct, the outward
{\it convective} heat flux acts like a reverse {\it diffusive} heat
flux, but one in which the thermal conductivity is itself proportional
to $(dT/dr)^{1/2}$.  Together with incomplete field-line insulation,
this is well-suited to act as an efficient regulatory mechanism
counteracting the otherwise large inward diffusive
transport of thermal energy.

\section{Discussion}

%
%
%
%Coulomb conduction in a dilute cluster plasma at temperatures of order
%$T>5\times 10^7\K$, in principle an intrinsic property of the plasma
%itself, is an extremely potent heating source.  Therein lies its
%difficulty: it is too efficient.  ``Unchecked'' thermal conduction
%would drive hot clusters to near isothermality, and this is as
%incompatible with observations as the cooling catastrophe scenario.
%Because of this difficulty, conduction is generally thought to be
%dramatically supressed, perhaps by tangled magnetic fields or by
%plasma instabilities.  However, as discussed in the introduction, even
%models where the conduction is assumed to be suppressed (by a global
%factor $f$) suffer fine-tuning and stability problems.  We argue that,
%in the cores of hot galaxy clusters, HBI-driven turbulent transport
%necessarily accompanies the conductive heat flux leading to convective
%cooling of the cluster core.

HBI-driven turbulence has two important consequences for heat
transport in hot cluster cores, the partial suppression of the
conductive heat flux due to field-line re-orientation (previously
stressed by PQ08) and an outwardly-directed convective heat flux
(highlighted in this work).  We hypothesize here that these two
effects keep a cooling core in marginal thermal stability, with the
(regulated) thermal conduction balancing radiative cooling.

Consider a hot cluster with an initially isothermal ICM, lacking any
embedded heat source such as an AGN.  Cooling in the core of the
cluster would set up an initially small temperature gradient and, for
a short time, the radiative loses would be easily compensated by
thermal conduction down this temperature gradient.  As time proceeds,
however, HBI induced field-line insulation and the convective heat
flux will act to suppress the conduction heat flux, leading to a
steepening temperature gradient.  As the center of the cluster
approaches a cooling catastrophe, the vigorous turbulence will comb
the field lines of the turbulent ICM preferentially in the radial
direction, opening up an enhanced channel for the inwards conductive
heat flux and quenching the runaway cooling.  We posit that dense ICM
cores will evolve towards a state of near static, marginal
``radiative-HBI'' stability, and identify this with the observed state
of the hot cluster cores.

Although this work appears to be the first to invoke turbulence as a
cooling mechanism, the suggestion that cluster cooling cores may be
turbulent is certainly not new (e.g., see Voigt \& Fabian 2004).
Indeed, the presence of turbulence is motivated by observational
studies of the rotation measure towards embedded radio sources (e.g.,
Vogt \& Ensslin 2005), spatially resolved ICM pressure fluctuations
(Schuecker et al. 2004), and X-ray resonant line scattering (e.g.,
Gastaldello \& Molendi 2004).  A direct observational link with our
proposal that HBI-driven turbulence is present in cluster cores is via
turbulent diffusion of metals throughout the core of the cluster.
Rebusco et al. (2005) estimate the iron injection rate into the
Perseus cluster as a function of radius, and find that significant
diffusion of the iron is required to explain the measured radial
profile of the iron abundance.  They estimate a diffusion constant
within the core of ${\cal D}\sim 2\times 10^{29}\,{\rm cm}^2\,{\rm
s}^{-1}$ and show that this can be reasonably accommodated by a
turbulent transport picture.  The source of turbulence is generally
attributed to the central AGN, either by direct stirring of the ICM by
AGN jets, or by AGN cosmic-ray driven convection (Chandran \& Rasera
2007).  Cluster-cluster mergers have also been considered as a source
of turbulence (Fujita, Matsumoto \& Wada 2004).

A distinguishing feature of our model is that it makes a direct
connection between the temperature gradient found in cluster cooling
cores and turbulent diffusion.  The inferred Rebusco et
al. (2005) diffusion constant is readily accommodated by
eq. (\ref{D}) if $\xi$ is order $0.1 r$.  Moreover, we also predict
efficient turbulent transport of metals within the cooling core, but
significantly {\em reduced} metal transport in the more isothermal
regions of the cluster.  Thus, the fact that the high metallicity
cores appear to be limited to the region where a temperature gradient
is maintained (e.g., Baldi et al. 2007; Snowden et al. 2008) is neatly
accounted for.

This scenario would require an associated convective heat
flux larger than is seen in numerical simulations of noncooling HBI
systems (Parrish \& Quataert 2008).  Estimates of the ratio of the convective
to local conductive fluxes based on equation (\ref{chf}) and the above
${\cal D}$ value are of order $10^{-2}$ to $10^{-1}$ for cluster cores.
The noncooling HBI simulations, by contrast, yield values of order $10^{-3}$.  
This will be an important point to reconcile in numerical
simulations of cooling HBI flows.  

Thus far, we have focused exclusively on hot clusters where unbridled
Spitzer conductivity is a more than sufficient heat source for the
cores.  How does our picture relate to clusters more generally?  Best
et al. (2007) have studied a large sample of clusters from the Sloan
Digital Sky Survey (SDSS) and gauged the role of radio-mode feedback
in these systems under the very reasonable assumption that the
relationship between radio-luminosity and mechanical power is not a
strong function of environment.  They find that, while embedded radio
galaxies can more than offset radiative cooling in galaxy groups and
small clusters, the AGN feedback hypothesis has significant energy
problems when one considers large/rich clusters.  They further show
that (when parameterized in terms of the virial velocity of the
cluster) the AGN feedback hypothesis becomes untenable just about
where the clusters are hot enough for thermal conduction to play a
significant role.  This coincidence is striking, and strong
circumstantial evidence that thermal conduction plays an important
role in regulating these hot clusters.

%
%
%Energetics aside, the case for AGN feedback as the answer to the
%cooling flow problem is far from complete.  First, even the deepest {\it
%Chandra} observations to date (e.g., the 1\,Msec study of the Perseus
%cluster; Fabian et al. 2006) have yet to find the direct signatures
%of ICM {\it heating} (e.g., sufficiently strong shocks).  Second, the
%implied regulation of the AGN power needed to balance radiative loses
%throughout the cooling core has yet to be explained.  Three-dimensional
%hydrodynamic models that assume AGN fueling from the central parts of the
%cooling flow have great difficulty in preventing a cooling catastrophe;
%ICM is effectively heated close to the (instantaneous) jet axis, but
%cooling occurs essentially unchecked away from the jet axis (Vernaleo \&
%Reynolds 2006).  Finally, the apparent similar behavior of cooling core
%temperature profiles across a wide range of mass scales, including an
%apparent temperature floor of $T_{\rm vir}/3$ where $T_{\rm vir}$ is the
%virial temperature (e.g., see Peterson \& Fabian 2006; Baldi et al. 2007)
%has yet to find an explanation within the AGN feedback hypothesis.

\section {Summary}

Cooling flows in galaxy clusters show remarkably regular temperature
profiles.  Naive cooling models, in which radiative losses are
replenished by advected heat inflow (i.e., classical cooling flow
models), fail badly, predicting an excessively large X-ray surface
brightness and large amounts of cold gas (or powerful star formation).
Direct heating from a central AGN is in principle a potent energy
source, but why this should result in regular temperature profiles and
temperature floors that scale with the virial temperature, is unclear.

Temperature gradients in the hot gas would produce a large inward heat
flux if Coulomb collisions alone were responsible for heat transport.
This process, which has in fact been extensively studied, does not
currently enjoy widespread support as an important generic heating
mechanism.  Heating by thermal conduction, when important, seems to be
overwhelming, and even if the match is good, thermal stability and
fine-tuning remains an issue.  For these reasons, researchers have
often looked to external sources of heat (e.g., the AGN) as the
resolution of the cooling flow problem.

We argue that the recent finding of Q08, that inward directing heat
flow in a dilute plasma is unstable and strongly self-regulating,
shows great promise as a mechanism to maintain regular temperature
profiles.  In this work, we have proposed that the combination of a
reverse convective heat flux and field-line insulation maintains
critical stability conditions corresponding to nearly static
stratified atmospheres.  In essence, cluster cores are in a state
of marginal stability, not to the HBI, but to the {\em radiative} HBI.
%is produced by the diffusive heat flux in cooling flows,
%along with field line insulation effects.  The convective heat flux
%(eq. \ref{chf}) maintains critical stability conditions corresponding
%to nearly static
%(in the sense of little or non inward mass flow)
%stratified atmospheres.  
Departures from this state should be stable: an increased diffusive
flux would produce a stronger reverse convective flux, and vice-versa.
A fully insulated cooling core would draw field lines inward, opening
a thermal path.  As a by-product of convection, we expect efficient
diffusion of metals in regions where a strong temperature gradient
exists.  This agrees with observations of enhanced ICM metallicities
over precisely the same region where the temperature gradient is
maintained, but the associated convective heat flux would be larger
than those inferred from nonradiative numerical simulations.
Radiative HBI simulations are clearly needed.  

The basic scenario outlined in this paper---that cluster cores are in
a state of marginal stability regulated by the radiative HBI---can
be tested numerically by direct simulation.  In a subsequent paper,
we will report the results of such a study.

\section*{Acknowledgements.}  
This work was supported by a Chaire d'Excellence award to SAB from the
French Ministry of Higher Education.
%and by NASA grants NNG04GK77G and NAG5-13288.
CSR thanks the Physics Department of the \'Ecole Normale
Sup\'erieure de Paris for its hospitality and support of a one month
visit in March 2008 during which this work was conducted.  CSR also
acknowledges support from the Chandra Theory and Modeling Program
under grant TM7-8009X.

\end{document}